\documentclass[aps,pra,twocolumn,showpacs,amsmath,amssymb,superscriptaddress]{revtex4-1}

\usepackage{tikz,hyperref}

\begin{document}

\title{Effective quantum memory Hamiltonian from local two-body interactions}

\author{Adrian Hutter}
\affiliation{Department of Physics, University of Basel, Klingelbergstrasse 82, CH-4056 Basel, Switzerland}
\author{Fabio L. Pedrocchi}
\affiliation{Institute for Quantum Information, RWTH Aachen University, D-52056 Aachen, Germany} 
\affiliation{Department of Physics, University of Basel, Klingelbergstrasse 82, CH-4056 Basel, Switzerland}
\author{James R. Wootton}
\affiliation{Department of Physics, University of Basel, Klingelbergstrasse 82, CH-4056 Basel, Switzerland}
\author{Daniel Loss}
\affiliation{Department of Physics, University of Basel, Klingelbergstrasse 82, CH-4056 Basel, Switzerland}

\newcommand{\m}[1]{\mathrm{#1}}
\newcommand{\nn}{\nonumber}
\newcommand{\kk}{_{\bf{k}}}
\newcommand{\mdag}{^{\dagger}}
\newcommand{\bos}{_{\m{bos}}}
\newcommand{\tr}{\mathop{\mathrm{Tr}}\nolimits}
\newcommand{\omg}{\omega}
\newcommand{\eps}{\epsilon}
\newcommand{\norm}[1]{\left\| #1\right \|}
\newcommand{\heff}{H_{\m{eff}}}
\newcommand{\id}{\mathbb{I}}
\newcommand{\vd}{V_{\m{d}}}
\newcommand{\vod}{V_{\m{od}}}
\newcommand{\mS}{{\mathcal S}}
\newcommand{\ra}{\rightarrow}

\newcommand{\px}{\sigma^x}
\newcommand{\py}{\sigma^y}
\newcommand{\pz}{\sigma^z}

\begin{abstract} 
In [Phys. Rev. A 88, 062313 (2013)] we proposed and studied a model for a self-correcting quantum memory in which the energetic cost for introducing a defect in the memory grows without bounds as a function of system size.
This positive behavior is due to attractive long-range  interactions mediated by a bosonic field to which the memory is coupled.
The crucial ingredients for the implementation of such a memory are the physical realization of the bosonic field as well as local five-body interactions between the stabilizer operators of the memory and the bosonic field.
Here, we show that both of these ingredients appear in a low-energy effective theory of a Hamiltonian that involves only two-body interactions between neighboring spins.
In particular, we consider the low-energy, long-wavelength excitations of an ordered Heisenberg ferromagnet (magnons) as a realization of the bosonic field. Furthermore, we present perturbative gadgets for generating the required five-spin operators.
Our Hamiltonian involving only local two-body interactions is thus expected to exhibit self-correcting properties as long as the noise affecting it is in the regime where the effective low-energy description remains valid.
\end{abstract}


\maketitle

\section{Introduction}
Kitaev's toric code \cite{Kitaev2003} serves as the simplest toy model of a quantum memory as well as being the archetypical example of a topological phase of matter.
Topological protection of Josephson junction qubits \cite{Gladchenko2009} as well as topological error correction \cite{Yao2012} have already been experimentally demonstrated.
Implementations of spin-lattice models with topologically ordered groundstates  using polar molecules stored in optical lattices \cite{Micheli2006} or laser-excited Rydberg atoms \cite{Weimer2010} have been proposed.

A challenge to any scalable implementation of topological protection of quantum information is the issue of thermal stability.
In its standard form, the toric code Hamiltonian requires a set of local, commuting four-qubit stabilizer operators $W=\left(\px\right)^{\otimes4}, \left(\pz\right)^{\otimes4}$.
Unfortunately, the ``bare'' toric code Hamiltonian $-A\sum W$ (with $A>0$) does not allow for thermally stable storage of quantum information \cite{Castelnovo2007,Nussinov2008,Alicki2009,Bravyi2009,Yoshida2011}. 
While performing a bit- or phase-flip on a single qubit (and thus creating two anyonics defects) has an energy cost $4A$, these defects can then propagate without any further energy cost and thus destroy the stored quantum information. 
As a consequence, the lifetime of the quantum information is \emph{independent} of the size of the memory.
Furthermore, interactions in nature are usually two-body, such that the four-body operators $W$ cannot be generated directly but have to emerge from an underlying structure of two-body interactions.
Since $W$ will then appear in high-order perturbation theory \cite{Kitaev2005,Koenig2010,Brell2011,Terhal2012}, the energy penalty $A$ will naturally be weak.

These negative results on the bare toric code have motivated the study of long-range interactions between the stabilizer operators $W$ 
as a means to suppress the creation and/or diffusion of anyons and thereby increase the lifetime of the stored quantum information \cite{Dennis2002,Hamma2009,Chesi2010,Pedrocchi2011,Beat2012,Hutter2012,Pedrocchi2013}.
Such long-range interactions can lead to quantum information lifetimes that grow polynomially or even exponentially with $L$ \cite{Hamma2009,Chesi2010,Beat2012,Hutter2012,Pedrocchi2013}.
It has been suggested that such memories are compatible with a recent proposal for fault-tolerant holonomic quantum computation based on adiabatic deformation of the system Hamiltonian \cite{Zheng2014}.

In order to mediate these interactions in a physically plausible way, all of these proposals require five-body operators of the form $W\otimes O$, 
where the operator $O$ allows to couple the stabilizer operator $W$ to an external field which mediates the interaction.
The external field can either be elementary (e.g.\ photons as discussed in Ref.~\onlinecite{Pedrocchi2011}) or emerge from the energetically low-lying excitations of a many-body system (e.g.\ phonons as discussed in Ref.~\onlinecite{Hamma2009}).
In Ref.~\onlinecite{Pedrocchi2013} we have studied in detail the case of bosons hopping in a cubic lattice, leading to a parabolic dispersion near the center of the Brillouin zone. 

In the present paper, we focus on the physical implementation of the quantum memory Hamiltonian proposed in Ref.~\onlinecite{Pedrocchi2013}. 
We start from a spin Hamiltonian involving only \emph{local two-body interactions} and study its low-energy theory via a perturbative Schrieffer-Wolff transformation. 
The bosonic field emerges from the energetically low-lying excitations of an ordered Heisenberg ferromagnet (magnons). For small wave-numbers, magnons indeed feature a parabolic dispersion.
Furthermore, we present perturbative gadgets for generating the five-body operators $W\otimes O$, describing interactions between the four-qubit stabilizer operators $W$ of the toric code and a spin operator $O$ of the ferromagnet.
Since the quantum-memory Hamiltonian of Ref.~\onlinecite{Pedrocchi2013} emerges as the effective low-energy/long-wavelength theory of our Hamiltonian with only local two-spin interactions,
this system is expected to exhibit self-correcting properties \emph{as long as this effective theory remains valid}.

The rest of this paper is organized as follows. We briefly review the main results of Ref.~\onlinecite{Pedrocchi2013} in Sec.~\ref{sec:previous} and discuss the relation to the present work.
In Sec.~\ref{sec:gadgets}, we present perturbative gadgets that allow effective five-body terms $W\otimes O$ to be obtained from local two-body interactions only.
In Sec.~\ref{sec:FM}, we show that using these five-body operators to couple the stabilizer operators $W$ to an ordered Heisenberg ferromagnet 
leads to a low-energy effective theory which coincides with the quantum memory Hamiltonian from Ref.~\onlinecite{Pedrocchi2013}. 
We can thus obtain an effective quantum memory Hamiltonian from a system (perturbative gadgets plus Heisenberg ferromagnet) with two-spin interactions only.
In Sec.~\ref{sec:validity} we study the regime in which this effective description is expected to be valid.
In Sec.~\ref{sec:backaction} we study the backaction of the coupling onto the ferromagnet and how to counteract it.
We conclude in Sec.~\ref{sec:conclusions}.

\section{Previous work}\label{sec:previous}

In Ref.~\onlinecite{Pedrocchi2013} we have studied the following model $H_{\m{int}}+H_b$ for a self-correcting quantum memory.
Consider a bosonic Hamiltonian
\begin{align}
 H_b = \sum\kk\omg\kk a\kk\mdag a\kk
\end{align}
with a dispersion, which in the low-${\bf k}$ limit is parabolic, $\omg\kk\approx D|{\bf k}|^2$.
The four-qubit stabilizer operators $W_p$ are arranged on a 2D array of size $L\times L$ and locally couple to the bosonic field,
\begin{align}\label{eq:Hint}
 H_{\m{int}} = A\sum_p W_pO_p\,,
\end{align}
where either $O_p=a_p\mdag a_p$ or $O_p=a_p + a_p\mdag$.
Here, $a_p$ and $a_p\mdag$ are Fourier transforms of the bosonic operators $a\kk$ and $a\kk\mdag$, $a_p=\frac{1}{\sqrt{N}}\sum\kk e^{i{\bf R}_p{\bf k}}a\kk$, where ${\bf R}_p$ is the spatial location of stabilizer $W_p$ and $N$ is the number of bosonic modes.
In other words, the operator $a_p$ ($a_p\mdag$) annihilates (creates) a boson at position ${\bf R}_p$.

\begin{figure}
	\centering
	\includegraphics[width=0.8\columnwidth]{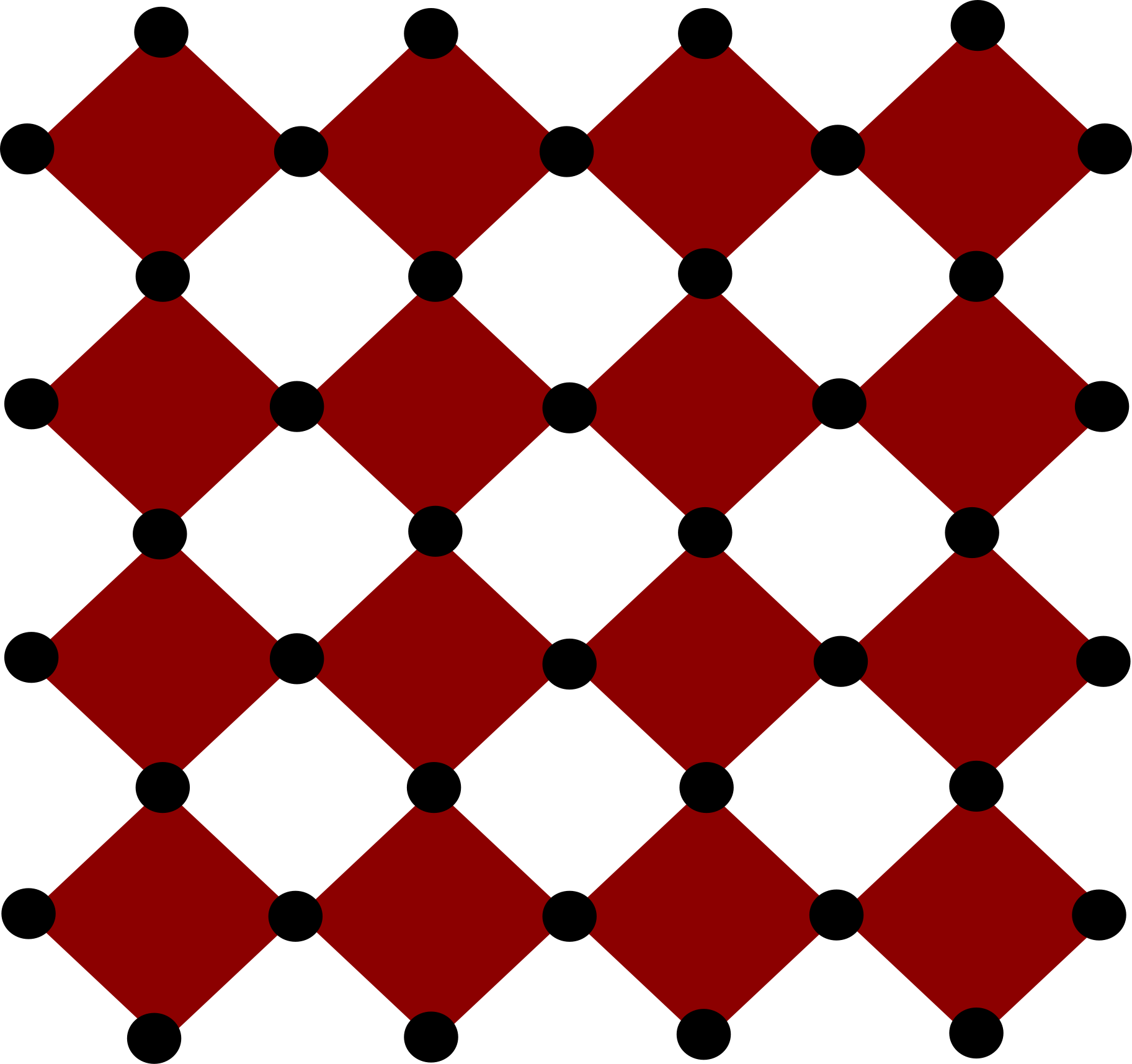}
	\caption{(Color online.) An excerpt of a toric code. Black dots are code qubits. Stabilizer operators $W_p$ involve operators acting on the four qubits around a white or a dark plaquette. 
	Operators acting on the four qubits around a white plaquette are of the form $W_p=(\px)^{\otimes4}$, while operators acting on the four qubits around a dark plaquette are of the form $W_p=(\pz)^{\otimes4}$.}
	\label{fig:code}
\end{figure}

The stabilizer operators (or stabilizers for short) $W_p$ are of the form $W_p=(\px)^{\otimes4}$ or $W_p=(\pz)^{\otimes4}$, see Fig.~\ref{fig:code} for an illustration.
All stabilizers commute with each other and have eigenvalues $\pm1$.

Under the assumption that the bosons are in thermal equilibrium, the bosonic field mediates long-ranged interactions between the stabilizer operators $W_p$.
Technically speaking, it is possible to integrate out the bosonic field (either exactly or perturbatively) and derive an effective Hamiltonian for the stabilizers. This Hamiltonian is of the form
\begin{align}\label{eq:Hstab}
 H_{\m{stab}} = \frac{1}{2}\sum_{p\neq p'}J_{pp'}W_pW_{p'}\,,
\end{align}
where $J_{pp'}$ describes a mediated attractive long-range interaction between the stabilizers.

More specifically, we have shown the following \cite{Pedrocchi2013}.
For $O_p=a_p\mdag a_p$, we have $J_{pp'}\sim|{\bf R}_p-{\bf R}_{p'}|^{-2}$, such that the energy cost for violating a stabilizer (``creating an anyon'') grows logarithmically with $L$ and the quantum memory lifetime grows polynomially with $L$.
If, on the other hand, $O_p=a_p + a_p\mdag$, we have $J_{pp'}\sim|{\bf R}_p-{\bf R}_{p'}|^{-1}$, such that the energy cost for creating an anyon grows linearly with $L$ and the lifetime does so exponentially.

\subsection{Relation to the present work}

The goal of this work is to show that the Hamiltonian $H_{\m{stab}}$ can be obtained as an effective low-energy Hamiltonian of a Hamiltonian that involves only nearest-neighbor two-spin interactions.
This Hamiltonian is then expected to exhibit self-correcting properties, given that the noise affecting the memory is such that the effective low-energy description remains valid.
We realize the bosonic field by a 3D ordered Heisenberg ferromagnet (FM) to which the stabilizers of the toric code couple locally.
The long-range interactions $J_{pp'}$ in Eq.~\eqref{eq:Hstab} are then mediated by massless excitations of the FM (Goldstone modes), so-called magnons.

Our model consists of a ``gadgetry'' and a ``ferromagnet'' part, \begin{align}H=H_G+H_F\,,\end{align} where $H_G=\sum_pH_p$ is a sum of identical gadget terms for each of the stabilizers.
The summands $H_p$ are sums of two-qubit terms, where one of the two qubits involved is part of the toric code and the other one is an auxiliary or ``mediator'' qubit.
By integrating out all mediator qubits, we obtain a first effective Hamiltonian
\begin{align}
 H' = (H_G)_{\m{eff}} + H_F\,,
\end{align}
describing four-qubit stabilizer operators locally coupled to the FM. Here, $(H_G)_{\m{eff}}$ is akin to $H_{\m{int}}$ in Eq.~\eqref{eq:Hint}.

The ordered Heisenberg ferromagnet described by $H_F$ can be mapped to a bath of bosons (magnons) by means of the well-known Holstein-Primakoff transformation.
Even beyond the one-magnon approximation (\textit{i.e.}, taking magnon-magnon interaction into account), the interactions between the stabilizers mediated by the ferromagnet can be described by a final effective Hamiltonian $\heff$ akin to $H_{\m{stab}}$ in Eq.~\eqref{eq:Hstab}. 
The effective interactions $J_{pp'}$ are thereby given by the static susceptibility of the FM.

\begin{figure}
	\centering
	\includegraphics[width=0.8\columnwidth]{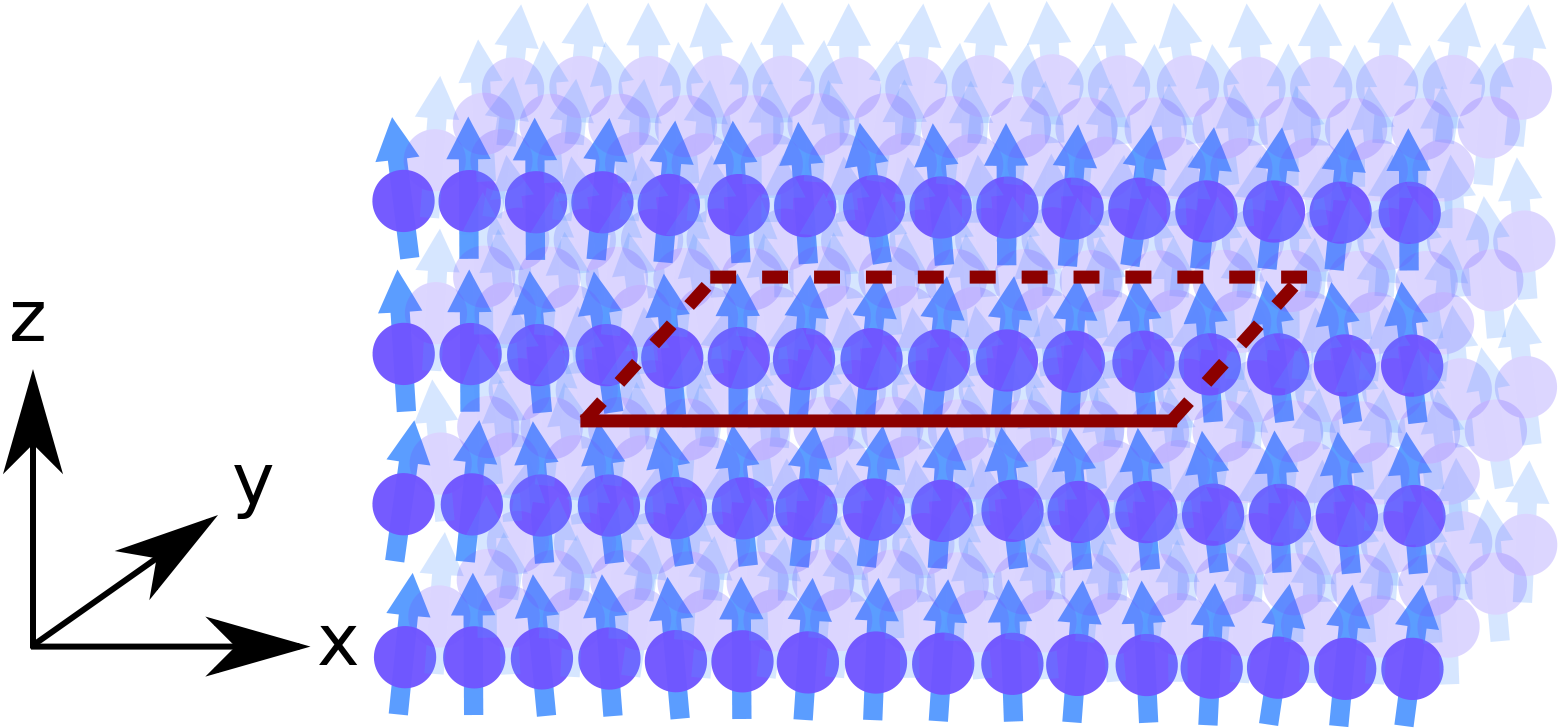}
	\caption{(Color online.) A 2D toric code (whose contoures, lying in an $xy$-plane, are sketched by red lines) is embedded in a 3D Heisenberg ferromagnet (blue) which is ordered in $z$-direction. 
	The stabilizer operators of the toric code $W_p$ (illustrated in Fig.~\ref{fig:code}) couple to the $x$-component of an adjacent spin ${\bf S}_p$ of the ferromagnet. The linear size of the planar toric code ($L$) is assumed to be much smaller than the one of the ferromagnet ($\Lambda$), {\it i.e.}, $L\ll \Lambda$.}
	\label{fig:FM}
\end{figure}

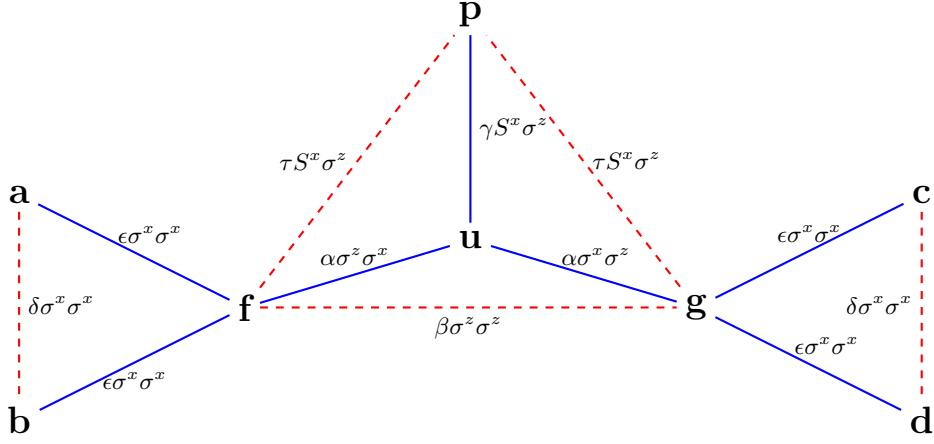
\begin{figure*}
\begin{tikzpicture}[xscale=6, yscale=1.5]
\node (b) at (0,0) {\Large\bf b};
\node (d) at (2,0) {\Large\bf d};
\node (a) at (0,2) {\Large\bf a};
\node (c) at (2,2) {\Large\bf c};
\node (f) at (0.5,1) {\Large\bf f};
\node (g) at (1.5,1) {\Large\bf g};
\node (u) at (1,1.6) {\Large\bf u};
\node (p) at (1,3.6) {\Large\bf p};

\draw [thick, blue] (a) -- (f) node [midway,above] {\color{black} $\quad\,\,\eps\px\px$};
\draw [thick, blue] (b) -- (f) node [midway,below] {\color{black} $\eps\px\px$};
\draw [thick, blue] (c) -- (g) node [midway,above] {\color{black} $\eps\px\px$};
\draw [thick, blue] (d) -- (g) node [midway,above] {\color{black} $\quad\,\,\,\eps\px\px$};
\draw [thick, blue] (f) -- (u) node [midway,above] {\color{black} $\alpha\pz\px$};
\draw [thick, blue] (g) -- (u) node [midway,above] {\color{black} $\quad\alpha\px\pz$};
\draw [thick, blue] (p) -- (u) node [midway,right] {\color{black} $\gamma S^x\pz$};

\draw [thick, dashed, red] (a) -- (b) node [midway,right] {\color{black} $\delta\px\px$};
\draw [thick, dashed, red] (c) -- (d) node [midway,left] {\color{black} $\delta\px\px$};
\draw [thick, dashed, red] (f) -- (g) node [midway,below] {\color{black} $\beta\pz\pz$};
\draw [thick, dashed, red] (f) -- (p) node [midway,left] {\color{black} $\tau S^x\pz$};
\draw [thick, dashed, red] (g) -- (p) node [midway,right] {\color{black} $\tau S^x\pz$};
\end{tikzpicture}
\caption{\label{fig:gadget} 
(Color online.) Auxiliary qubits $f$, $g$, and $u$ mediate a five-body interaction $W_p\otimes S^x_p$ between qubits $a$, $b$, $c$, and $d$, and the operator $S^x_p$ (spin of the FM). 
This interaction emerges from local two-body interactions only. 
The excited state of the auxiliary qubits is penalized by an energy $\Delta$, which is the dominant energy scale in the system.
The interactions which are indicated by solid lines produce the actual five-body interaction.
Interactions which are indicated by dashed lines allow one to tune the strength of two-, three-, and four-body terms without changing the strength of the five-body term.
Choosing $\delta$ and $\tau$ appropriately allows one to counter undesired two- and three-body terms.
Finally, the parameter $\beta$ allows one to tune the strength of the four-qubit interaction $W_p$ independently of the five-body interaction $W_p\otimes O_p$.}
\end{figure*}

\section{Perturbative gadgets for five-body operators}\label{sec:gadgets}

Given some Hamiltonian with only local two-body interactions, one is often interested in an \emph{effective} Hamiltonian that describes the low-energy dynamics of the system. This is achieved by ``integrating out'' the high-energy degrees of freedom.
The effective low-energy Hamiltonian then often features a higher complexity than the original one. 
This lead to the idea of \emph{perturbative gadgets} \cite{Kempe2006,Jordan2008,Oliveira2008}, which allows one to systematically construct Hamiltonians with local two-qubit interactions that yield some desired effective (low-energy) few-qubit Hamiltonian.
In particular, Ref.~\onlinecite{Oliveira2008} showed that any local Hamiltonian can be generated from some $2$-local Hamiltonian. 

The Schrieffer-Wolff (SW) transformation \cite{SW1966,Bravyi2011} (see Appendix~\ref{app:SW} for a technical summary) provides a natural framework for obtaining such effective terms.
Ref.~\onlinecite{Bravyi2008} combined the gadgets of Ref.~\onlinecite{Oliveira2008} with the SW method to discuss the simulation of local many-body Hamiltonians by use of $2$-local interactions.
The schemes we propose in order to generate the five-body terms $AW_p\otimes O_p$ shall be analyzed by means of a SW transformation but are simpler than had they been constructed with the perturbative gadgets described in Refs.~\onlinecite{Oliveira2008,Bravyi2008}.

Note that while all of the proposals in Refs.~\onlinecite{Dennis2002,Hamma2009,Chesi2010,Pedrocchi2011,Beat2012,Hutter2012,Pedrocchi2013} require five-body terms of the form $AW_p\otimes O_p$, 
other quantum memory proposals are not based on coupling stabilizer operators to external fields. 
However, these alternate proposals in fact involve interactions between more than five qubits.
Specifically, the 3-dimensional toric code with ``welding'' \cite{Michnicki2012} requires six-qubit operators, while the 4-dimensional toric code \cite{Dennis2002} and Haah's cubic code \cite{Haah2011a,Haah2011b,Bravyi2013} require eight-qubit operators.
These models can thus only be realized with gadgets that are even more involved than the ones discussed in the following.

We introduce two sets of spins, namely ${\bf S}_{j}$ for the spins of the  3D  FM located at site $j$ of a cubic lattice and $(\px_i,\py_i,\pz_i)^T$ for the physical spins-$1/2$ (qubits) of the 2D toric code. Both spins satisfy the usual commutation relations. 
The code qubits are arranged on a quadratic lattice with periodic boundary conditions.
The four-qubit stabilizer operators $W_p$ are of the form $\px_a\px_b\px_c\px_d$ or $\pz_a\pz_b\pz_c\pz_d$,
where the operators act on the four qubits around one plaquette of the lattice, as illustrated in Fig.~\ref{fig:code}.

We seek to construct effective terms of the form $W_p\otimes S_p^x$, where for the moment we consider $W_p=\px_a\px_b\px_c\px_d$.
Here, ${\bf S}_p$ is the spin of the FM adjacent to the spins of the stabilizer $W_p$, see Fig.~\ref{fig:FM}.
Let the summands in the gadget Hamiltonian $H_G=\sum_pH_p$ be given by   
\begin{align}\label{eq:gadget}
 H_p = &\,\, -\frac{\Delta}{2}\pz_f - \frac{\Delta}{2}\pz_g - \frac{\Delta}{2}\pz_u \nn\\
& + \gamma S_p^x\otimes \pz_u + \tau S_p^x \otimes(\pz_f+\pz_g) \nn\\
& + \eps \px_f\otimes(\px_a+\px_b) + \eps \px_g\otimes(\px_c+\px_d) \nn\\
& + \alpha\px_u\otimes(\pz_f+\pz_g) + \beta \pz_f\otimes \pz_g \nn\\
& + \delta\px_a\otimes\px_b + \delta \px_c\otimes\px_d 
\end{align}
with $\Delta$ being larger than (the absolute value of) all other energies (see Fig.~(\ref{fig:gadget}) for an illustration).

We now apply the SW method to successively integrate out the mediator qubits $f$, $g$, and $u$ and calculate for each mediator qubit the resulting terms up to third order.
Consider a mediator qubit $r$ with energy splitting $\Delta$, $H_0=-\frac{\Delta}{2}\pz_r$, and a perturbation $V=\vd+\vod$.
With $\vd=\pz_r\otimes\tilde{V}_{\m{d}}$, $\vod=\px_r\otimes\tilde{V}_{\m{od}}$, and $[\tilde{V}_{\m{d}},\tilde{V}_{\m{od}}]=0$, 
we obtain after integrating out the auxiliary qubit $r$ as described in Appendix~\ref{app:SW} 
\begin{align}\label{eq:SWfinal}
 \heff = -\frac{\Delta}{2} + \tilde{V}_{\m{d}} - \frac{1}{\Delta}\tilde{V}_{\m{od}}^2 - \frac{2}{\Delta^2}\tilde{V}_{\m{od}}^2\tilde{V}_{\m{d}} + \ldots
\end{align}
The unitaries applied during the SW procedure to integrate out qubits $f$, $g$, and $u$ (cf.\ Appendix~\ref{app:SW}) do not commute and higher order terms will thus depend on the order in which this three qubits are integrated out.
However, up to the orders stated below, the effective terms are independent of this ordering.

For the sake of a shorter notation, let $R_p:=\px_a\px_b+\px_c\px_d$ and $\xi:=\frac{2\eps}{\Delta}$.
We neglect terms in the interaction strengths which are smaller by at least a factor $\xi^2$ than the mentioned terms.
After straightforwardly applying Eq.~\eqref{eq:SWfinal} in Appendix~\ref{app:SW} to the mediator qubits $f$, $g$, and $u$, we find
\begin{align}\label{eq:heffFinal}
(H_p)_{\m{eff}} &= \left[\gamma + 2\tau - 8\frac{\gamma\alpha^2}{\Delta}\right]S_p^x - \xi^2\left[\tau - 8\frac{\alpha^2\gamma}{\Delta^2}\right]R_p\otimes S_p^x \nn\\
&\quad + \left[\delta - \xi^2\left(\frac{\Delta}{2} -\beta + 4\frac{\alpha^2}{\Delta}\right)\right]R_p + \xi^4\left[\beta-2\frac{\alpha^2}{\Delta}\right]W_p \nn\\
&\quad - 4\xi^4\frac{\alpha^2\gamma}{\Delta^2}W_p\otimes S_p^x + \m{const}\ .
\end{align}
The parameter $\beta$ allows one to tune the strength of the plaquette term $W_p$ without affecting the strength of the five-body operator $W_p\otimes S_p^x$.
The interaction strength of the undesired operators $R_p$ and $R_p\otimes S_p^x$ can be tuned to zero through appropriate choice of the parameters $\delta\simeq\frac{2\eps^2}{\Delta}$ and $\tau\simeq8\frac{\alpha^2\gamma}{\Delta^2}$, respectively.
We obtain an undesired one-body term $\simeq(\gamma+2\tau)S_p^x$ which can be countered by an appropriate local field.
Finally, the strength of our desired five-body term $W_p\otimes S_p^x$ is to leading order given by $A:=-64\frac{\eps^4\alpha^2\gamma}{\Delta^6}$. 
If we use the operator $W_p\otimes S_p^x$ to couple the plaquette $W_p$ to external fields, effective two-plaquette interactions mediated by the field will be of even order in $A$.
The sign of $\alpha$, $\gamma$, and $\eps$ is thus irrelevant. 

Relabeling $x\rightarrow z\rightarrow y\rightarrow x$ on all qubits, such that the commutation relations of the spin operators are preserved, we can obtain operators of the form $\left(\pz\right)^{\otimes4}\otimes S_p^x$ rather than $\left(\px\right)^{\otimes4}\otimes S_p^x$.
In conclusion, integrating out the mediator qubits in each gadget leads to an effective coupling Hamiltonian
\begin{align}\label{eq:Heff}
 (H_G)_{\m{eff}}=\sum_pA W_p\otimes S_p^x\,,
\end{align}
which is of the same form as Eq.~\eqref{eq:Hint}.

Note that if we let $\alpha,\gamma,\tau\rightarrow0$ in Eq.~(\ref{eq:gadget}), we find the simpler effective Hamiltonian
\begin{align}
 (H_p)_{\m{eff}} = \left[\delta - 2\frac{\eps^2}{\Delta} - 4\frac{\beta\eps^2}{\Delta^2}\right]R_p + 16\frac{\beta\eps^4}{\Delta^4}W_p\ .
\end{align}
We can thus obtain stabilizer operators $W_p$ as they appear in Kitaev's toric code \cite{Kitaev2003} as effective terms using only two auxiliary qubits, nearest neighbor Ising interactions and single-qubit energy splittings.
This adds a particularly simple gadget to the list of proposals for perturbatively generating Kitaev's toric code \cite{Kitaev2005,Koenig2010,Brell2011,Terhal2012}.

We have discussed how the strength of the undesired terms $R_p$ and $R_p\otimes S^x_p$ can be tuned to zero by appropriate choice of the interaction strengths $\delta$ and $\tau$.
Of course, assuming that the strength of these interactions vanishes exactly is unphysical. Weak terms acting as $R_p$ are no threat to the toric code, as it is inherently stable against such local perturbations \cite{Kitaev2003,Bravyi2010}.
However, as the terms $R_p\otimes S^x_p$ themselves couple to the gapless magnon field, one might fear that their combined non-local interaction may destabilize the toric code groundstate.
We have qualitatively discussed the effect of such non-locally coupled perturbations on the stability of the toric code in Ref.~\onlinecite{Pedrocchi2013} 
and argued that they pose no threat to toric code stabilized by the FM as long as the strength of the accidental terms $R_p\otimes S^x_p$ is sufficiently smaller than the engineered coupling strength $A$.

\section{Effective long-range interactions mediated by the ferromagnet}\label{sec:FM}

After integrating out all mediator qubits in the perturbative gadgets, we arrived at a first effective Hamiltonian
\begin{align}\label{eq:Hprime}
 H'=(H_G)_{\m{eff}}+H_F=A\sum_pW_p S_p^x+H_F 
\end{align}
describing four-qubit stabilizer operators of the toric code coupled to the FM. 
Let us now study the interactions between the effective stabilizer operators $W_p$ which are mediated by the ordered Heisenberg FM $H_F$, to which they are coupled over the operators $S_p^x$.

The Hamiltonian of the 3D Heisenberg FM is given by $H_{\m{F}}=-J\sum_{\langle i,j\rangle}{\bf S}_{i}\cdot{\bf S}_{j}+h_{z}\sum_{i}S_{i}^{z}$,  where  $J>0$ is the exchange constant and  the sum is restricted to nearest-neighbor lattice sites. 
The FM is of linear size $\Lambda$, which is much larger than the linear size of the toric code, $\Lambda\gg L$.
The FM is assumed to be below the Curie temperature and the spins ordered  along the $z$-direction. To break the symmetry of the FM, a small magnetic field $h_{z}$ in $z$-direction is applied.
This field also stabilizes the FM against the effective longitudinal magnetic field produced by coupling the stabilizer operators of the toric code to the $x$-component of adjacent FM spins (see below). 
Although $H_F$ is three-dimensional, we point out for the sake of clarity, that the actual quantum memory is the two-dimensional toric code. 
The presence of the 3D system is necessary to mediate long-range interactions between the stabilizers. However, the place where the logical qubits are stored is the two-dimensional toric code.

For $A\ll J$, we make use of a perturbative second-order Schrieffer-Wolff transformation \cite{SW1966,Bravyi2011,Braunecker2008} to derive the effective plaquette-plaquette interaction (see Appendix~\ref{app:SWtransverse}) given by
\begin{equation}\label{eq:effectivecoupling}
 \heff=\frac{1}{2}\sum_{p\neq p'}J_{pp'}W_{p}W_{p'}\ ,
\end{equation}
where the  coupling is $J_{pp'}=-A^2\chi_{xx}({\bf R}_p-{\bf R}_{p'})$ and $\chi_{\alpha\beta}({\bf r})$ is the static spin susceptibility of the FM.
This effective Hamiltonian is of the same form as in Eq.~\eqref{eq:Hstab} and, as we will discuss now, the mediated interaction strength $J_{pp'}$, too, is of the same form.

The real space static susceptibility $\chi_{\alpha\beta}(\bf r)$ is defined as the Fourier transform of
\begin{equation}\label{eq:susceptibility}
\chi_{\alpha\beta}({\bf q},\omega)=i\lim\limits_{\eta\rightarrow0^{+}}\int_{0}^{\infty}\m{d}t\,e^{(i\omega-\eta) t}\left\langle\left[S_{{\bf q}}^{\alpha}(t),S_{-{\bf q}}^{\beta}\right]\right\rangle\ ,
\end{equation}
for $\omega=0$, where $\left\langle\ldots\right\rangle$ denotes thermal equilibrium expectation values of the ${\bf S}$-spins at temperature $T$. 
The Fourier components are defined as $S_{\bf q}^{\alpha}=\frac{1}{\sqrt{N_{s}}}\sum_{i}e^{-i{\bf q}\cdot{\bf R}_i}S_i^{\alpha}$, where $N_{s}=\Lambda^3$ is the number of spins in the FM, and ${\bf R}_{i}$ is a 3D vector pointing to the site of spin ${\bf S}_{i}$ of the FM. 

It is not necessary to explicitly calculate the spin susceptibility in the ferromagnetically ordered state to understand its general behavior at large distances (or small ${\bf q}$) \cite{Forster}. Indeed, for $h_{z}=0$, the spontaneous $SO(3)$ symmetry breaking of the state with finite magnetization pointing along the $z$-axis, implies the presence of low-frequency Goldstone modes (called magnons in this context) and long-range correlations, \textit{i.e.}, the $xx$- (and $yy$-) susceptibility has to diverge for ${\bf q}\rightarrow{\bf 0}$ and takes the following generic form in the hydrodynamic regime (low-energy and long wavelength regime) \cite{Forster}
\begin{equation}\label{eq:susceptibility}
\chi_{xx}({\bf q},\omega=0)=\frac{M^2}{\rho\vert{\bf q}\vert^2}\,\,\,\, \,\,\, \m{for}\,\,{\bf q}\rightarrow{\bf 0}\,.
\end{equation}
Here, $\rho>0$ is the stiffness constant of the FM and $M=\langle s^z\rangle$ is the magnetization density with $s^z=\frac{1}{N_{s}}\sum_{i}S_{i}^z$.
The divergence at ${\bf q}\rightarrow {\bf 0}$ in Eq.~(\ref{eq:susceptibility}) is directly connected with the broken symmetry of the ground state: 
starting from a ferromagnetic state aligned along the $z$-direction, the slightest $x$-magnetic field is able to rotate and align all spins in $x$-direction and thus the response to an external magnetic field indeed diverges at ${\bf q}\rightarrow {\bf 0}$.

Eq.~(\ref{eq:susceptibility}) is the expression for the spin susceptibility in the continuum approximation (lattice constant $a$ going formally to zero). 
To be valid this approximation does not require that the number of spins goes to infinity, but rather that the distance between neighboring spins is much smaller than the distances we are interested in. 
Since we are concerned with the long-distance physics of our model on the scale of $L$, this approximation is justified and simply requires $a/L\ll 1$. 
In this limit, both the lattice constants of the ferromagnet and of the toric code are taken to zero such that a single plaquette remains coupled to a single FM spin.

Below we give an explicit expression for the stiffness $\rho$ in the one-magnon approximation. The presence of the symmetry-breaking magnetic field $h_{z}$ introduces a gap in the magnon spectrum and thus a mass term in the susceptibility, \textit{i.e.}, $\chi_{xx}({\bf q},\omega=0)=\frac{M^2}{\rho\vert{\bf q}\vert^2+Sh_{z}}\,\m{for}\,{\bf q}\rightarrow{\bf 0}$.
The real space static susceptibility now follows by Fourier transformation which leads to 
$\chi_{xx}({\bf r})=\frac{M^2}{\rho}\frac{1}{4\pi\vert{\bf r}\vert}e^{-\vert{\bf r}\vert/L_{h}}$, with magnetic length $L_{h}=\sqrt{R/Sh_{z}}$.
Consequently, Eq.~(\ref{eq:effectivecoupling}) describes a stabilizer Hamiltonian with plaquette-plaquette interactions given by a Yukawa-like potential,
\begin{eqnarray}\label{eq:coupling_stiffness}
J_{pp'}=-\frac{A^2 M^2}{4\pi \rho}\frac{e^{-\vert{\bf R}_p-{\bf R}_{p'}\vert/L_{h}}}{\vert{\bf R}_p-{\bf R}_{p'}\vert}\,.
\end{eqnarray}
Since, again, $\rho>0$  (see also below), the interaction between stabilizer operators $W_{{\bf p}}$ is {\it attractive}.

For the sake of illustration we calculate $\rho$  in the one-magnon (harmonic) approximation by making use of  the Holstein-Primakoff transformation
\begin{equation}\label{eq:HP}
S_{i}^{z}=-S+{\hat n}_i\,,\,\,\,\,S_{i}^{-}=a_i^{\dagger}\sqrt{2S-{\hat n}_{i}} \,,\,\,\, S_{i}^{+}=(S_{i}^{-})^{\dagger},
\end{equation}
in the formal limit, where the occupation ${\hat n}_i=a_{i}^{\dagger}a_{i}$ is much smaller than $2S$ \cite{Nolting}. 
Here, $a_{i}$ and $a_{i}^{\dagger}$ satisfy bosonic commutation relations and the associated quasi-particles are the well-known magnons (or spin wave excitations). 
In Fourier space, we get $H_{F}\approx\sum_{\bf q}(\omega_{\bf q}+h_z)a_{\bf q}^{\dagger}a_{\bf q}$, up to some irrelevant constant, with magnon dispersion $\omega_{\bf q}=4JS[3-(\cos(q_x)+\cos(q_y)+\cos(q_z))]$, where 
$a_{\bf q}=\frac{1}{\sqrt{N_{s}}}\sum_{i}e^{-i{\bf q}\cdot{\bf R_i}}a_i$, with $N_s$ the number of FM spins. Inserting  Eq.~(\ref{eq:HP}) into Eq.~(\ref{eq:susceptibility}) and using a small ${\bf q}$ expansion leads to
$\chi_{xx}^{(0)}({\bf q},\omega=0)=\frac{S}{2JS\vert{\bf q}\vert^2+h_{z}}$, which allows us to identify the stiffness in lowest order $\rho^{(0)}=2JS^2$ since here $M^{(0)}=-S$. We thus obtain $\chi_{xx}^{(0)}({\bf r})=\frac{1}{8\pi J\vert{\bf r}\vert}e^{-\vert{\bf r}\vert/L_{h}}$ and  from this the approximate plaquette coupling 
\begin{equation}\label{eq:finalcoupling}
J_{pp'}^{(0)}
=-\frac{A^2}{8\pi J}\frac{e^{-\vert{\bf R}_p-{\bf R}_{p'}\vert/L_{h}}}{\vert{\bf R}_p-{\bf R}_{p'}\vert} ,
\end{equation}
which is explicitly attractive since $J>0$. We emphasize  that Eq.~(\ref{eq:finalcoupling}) is the one-magnon approximation of Eq.~(\ref{eq:coupling_stiffness}). 
The sole effect of both temperature and magnon-magnon interactions is to renormalize the coefficients of the interaction (\ref{eq:finalcoupling}), {\it i.e.}, $(M^{(0)})^2/R^{(0)}\rightarrow M^2/R$ \cite{Forster}, while the form of the potential is not affected. 
Note that the dimensionality of the  3D FM  is  critical since Heisenberg FMs in lower dimensions  do not order at $T>0$ \cite{Mermin1966}.

If $h_z$ is small enough such that $L_{h}\gg L$, the transverse susceptibility of the FM and hence the mediated interaction $J_{pp'}$ in Eq.~\eqref{eq:coupling_stiffness} decay like $\vert{\bf R}_p-{\bf R}_{p'}\vert^{-1}$ on the length-scale $L$ of the toric code.
The same decay of $J_{pp'}$ was reported in Sec.~\ref{sec:previous} for the case of coupling to the bosonic operator $O_p=a_p+a_p\mdag$.
This is of course no surprise, since the spin operator $S^x_p$ takes in the Holstein-Primakoff picture and in the limit $|\langle{\hat n}_i\rangle|\ll 2S$ indeed the form $\sqrt{2S}(a_p+a_p\mdag)$.

The external magnetic field $h_{z}$ is necessary for stabilizing the magnetization of the FM, keeping it along the $z$-direction. 
Indeed, the only condition which needs to be satisfied for the stability of the global magnetization of the FM is that the Zeeman energy $E_{z}=h_{z}S\Lambda^3$ due to the $h_{z}$ field remains much larger than the Zeeman energy $E_{x}=ASL^2$ due to the toric code. 
As a specific example, one can make the following scaling choice satisfying all constraints: $h_{z}\propto1/L^4$ and $\Lambda\propto L^3$, which satisfy $L_{h}\propto L^2\gg L$ and $E_{z}/E_{x}\propto L^3\gg 1$.  
Under these conditions, it is clear that the total magnetization will not be affected by the presence of the memory and the FM spins will not rotate into the $x$-direction{\it on average}. 
This is in agreement with a Metropolis simulation of the classical Heisenberg FM, see Fig.~\ref{fig:numerics}. However, we show below that backaction effects become eventually important for the FM spins {\it close} to the memory. 

 \section{Validity of the effective theory}\label{sec:validity}

Here we analyze in detail  the conditions of validity of our effective theory.

As we discussed in Ref.~\onlinecite{Pedrocchi2013}, the thermal stability of the toric code protected by the effective Hamiltonian $\heff$ in Eq.~\eqref{eq:effectivecoupling} is due to the fact 
that the cost for inverting a stabilizer $W_p$ (creating an anyon) grows without bounds as a function of $L$.
Indeed, assume that initially all stabilizers have a $+1$ eigenvalue ($W_p\equiv+1$). Then, the energy cost for inverting one of them is given by
\begin{align}
 \mu(L) = \sum_p2|J_{pp'}|\sim \frac{A^2M^2}{\rho}L\,,
\end{align}
where we used Eq.~\eqref{eq:coupling_stiffness} and the assumption $L_h\gg L$.
This seems to be in contradiction to the fact that our original Hamiltonian $H=H_G+H_F$ is a sum of bounded local terms, such that the energy cost for a local change certainly is bounded and does not grow with system size.
However, in the case where the effective Hamiltonian (\ref{eq:effectivecoupling}) emerges from a system with two-body interactions, it describes the system correctly in the \textit{long-wavelength} and \textit{low-energy limit} only. 

We first note that the Schrieffer-Wolff transformation employed in the derivation of $\heff$, followed by tracing out the degrees of freedom of the mediator qubits and the ferromagnet is not a unitary operation. 
Hence, the spectra of $H$ and $\heff$ are, in general, not identical. They match only in the low-energy sector where $H_{\m{eff}}$ leads to bounded results. 
Indeed, when the quantum memory is in contact with a thermal heat bath at inverse temperature $\beta$, the expected anyon density (fraction of stabilizer operators $W_p$ with a $-1$ value) is approximately $1/(e^{\beta\mu(L)}+1)$ 
(cf.\ Sec.~IV in Ref.~\onlinecite{Pedrocchi2013} for a more careful discussion, taking inter-anyonic interactions into account).
The thermal energy of the code system is thus 
\begin{equation}\label{eq:energy}
\langle H_{\m{eff}}\rangle\approx L^2\frac{\mu(L)}{e^{\beta\mu(L)}+1}\,,
\end{equation}
which vanishes exponentially as $L\ra\infty$.
The thermal density of anyons is self-consistently suppressed by the effective anyon chemical potential $\mu(L)$.
The total thermal energy $\langle H_{\m{eff}}\rangle$ remains thus finite and small even for large $L$. 
The fact that $\mu(L)$ diverges as a function of $L$ therefore does not invalidate our effective theory when its full effects are taken into account. 

The effective description breaks down in the high-temperature limit, $\beta\mu(L) \ll 1$, which corresponds to populating high-energetic states with finite probability, \textit{i.e.}, the anyon density becomes of order unity. 
Indeed, in this limit, we get $\langle H_{\m{eff}}\rangle\sim \mu(L)L^2$, which is clearly in contradiction with a finite upper bound on the energy density $\langle H_{\m{eff}}\rangle/L^2$.
This breakdown of the low-energy effective theory is of course not surprising, since in this regime the perturbative gadgets no longer work and the magnon expansion for the FM becomes invalid.
(For $T>\Delta,J$, the fraction of excited mediator qubits and the magnon occupation numbers become of order unity.)

Thus, our effective long-wave length and low-energy description is self-consistent for sufficiently low temperatures $T$ and sufficiently large code sizes $L$. 
This is similar to e.g.\ the harmonic approximation of crystal vibrations described by phonons. 
The Hamiltonian $H_{\m{phonon}}=\sum_{\bf q}D\vert {\bf q}\vert n_{\bf q}$ is only valid in the low-energy regime, and high-energy (large $\bf q$) excitations are self-consistently suppressed by the Bose-Einstein factor 
$\langle n_{\bf q} \rangle =1/(e^{\beta D \vert{\bf q}\vert}-1)$. 
 
The effective Hamiltonian (\ref{eq:effectivecoupling}) is not suited to describe the high-energy part of the spectrum, where the anyon density is of order unity. 
High-energy excitations are produced when anyons are created non-adiabatically, forcing the mediator qubits and the FM to leave their local equilibrium. 
In such a scenario the gadgets and the FM have no time to react and do not penalize the creation of anyons. 
In fact, the energy cost to create an anyon ``instantaneously'' is bounded by a finite constant as argued above. 
The fast creation of an anyon produces a bunch of high-energy and short-wavelength excitations in the gadgets/FM and these kinds of processes are not described by our effective theory. 

In the following, we provide analytical expressions for the regime of validity of our effective theory.
In the derivation of Hamiltonian (\ref{eq:effectivecoupling}) we explicitly assumed that mediator qubits reside in their groundstate and the FM is locally aligned along $z$-direction. 
This has to remain valid when thermal anyons are produced. 
So besides the requirement that the coupling $B$ to the external bath is small, \textit{i.e.}, $|B|\ll A$, we work in the {\it adiabatic regime} where the external bath creates errors in the code on a timescale much longer than $1/A$, or in other words, when the error rate is much smaller than $A$. 
For instance, modeling the coupling between code and bath by a generic spin-boson model \cite{DiVincenzo2005},
the error rate $\gamma(\omega)$ describing processes in which an energy $\omg$ is transferred from a code qubit to the bath takes the following form \cite{Chesi2010}
\begin{equation}
\gamma(\omega)=\kappa_{n}\left\vert \frac{\omega^n}{1-e^{-\beta\omega}}\right\vert e^{-\omega/\omega_{c}}\,,
\end{equation}
where $\omega_{c}$ is an arbitrary cut-off and $\kappa_{n}$ contains the coupling $B$ to the external bath (in Born approximation, $\kappa_{n}\propto B^2$). For $n=1$ the bath is called Ohmic, while it is called super-Ohmic for $n\geq2$. 
The adiabaticity condition then simply reads $\gamma(-A)\ll A$.  In this case, the mediator qubits and the FM have enough time to adapt to the perturbation generated by an error and stay respectively in an unexcited state or locally aligned along $z$-direction. 
In this regime, trying to flip a single qubit of the code will ``drag along'' a large number of other spins, leading to a large effective energy penalty.
To conclude, in such a scenario the low-energy description $\heff$ of the two-spin Hamiltonian $H$, involving long-range interactions between four-qubit operators, remains valid.

\section{Backaction effects onto the ferromagnet}\label{sec:backaction}

\begin{figure*}
	\centering
	\setlength{\unitlength}{0.8\columnwidth}
	\begin{picture}(1,1.2)
	  \put(-0.5,-0.05){\includegraphics[width=0.7\textwidth]{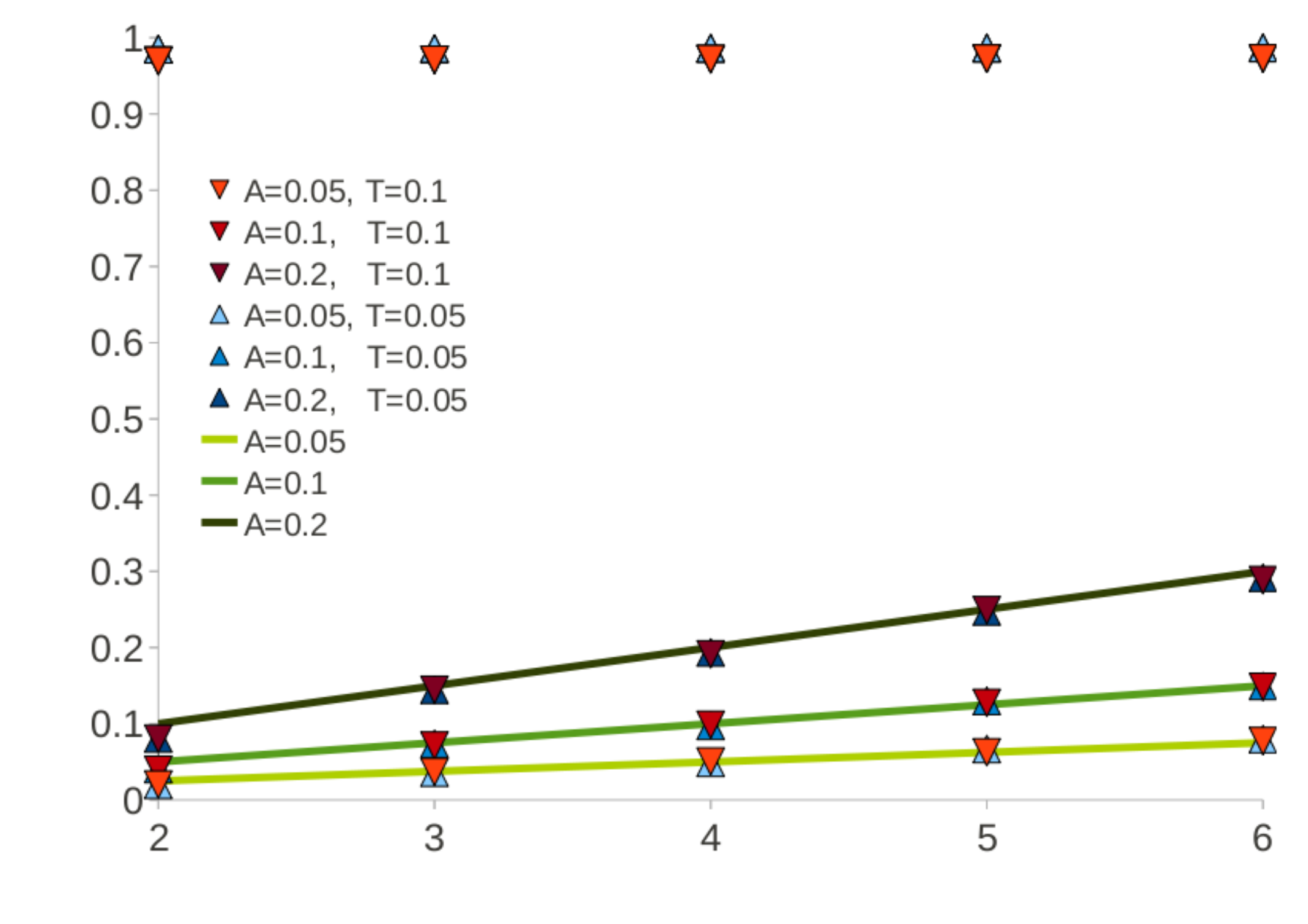}}
	  \put(1.00,0.46){\large $\langle S_{i}^{x}\rangle$}
	  \put(0.75,1.05){\large $\frac{1}{N_{s}}\sum_{i}\langle S_{i}^z\rangle$}
	  \put(1.05,0.02){\large $L$}
	\end{picture}

\caption{(Color online.) A graph of $\langle S_{i}^{x}\rangle$ (in the middle of the code) against $L$ for the classical Heisenberg FM with $J=1$. The data was obtained numerically by using the Metropolis algorithm. 
The magnetic length is $L_h=L^2$ and the FM size is $\Lambda=2 L_h$. The data shows agreement to the relation $S_{i}^{x}(t\rightarrow\infty) \propto LA /J$, obtained from Eq.~(\ref{eq:xSpin}). 
On the same graph we plot the total  $z$-magnetization $\frac{1}{N_{s}}\sum_{i}\langle S_{i}^{z}\rangle$ against $L$ demonstrating that the backaction is only a localized effect. The scaling chosen here is different than the one in the main text. 
This different choice is only motivated by the difficulty to simulate the classical Heisenberg ferromagnet with a large number of spins. 
This does not alter the analysis since the chosen scaling satisfies the necessary requirements $E_{z}\gg E_{x}$ and $L_{h}\gg L$.}
	\label{fig:numerics}
\end{figure*}
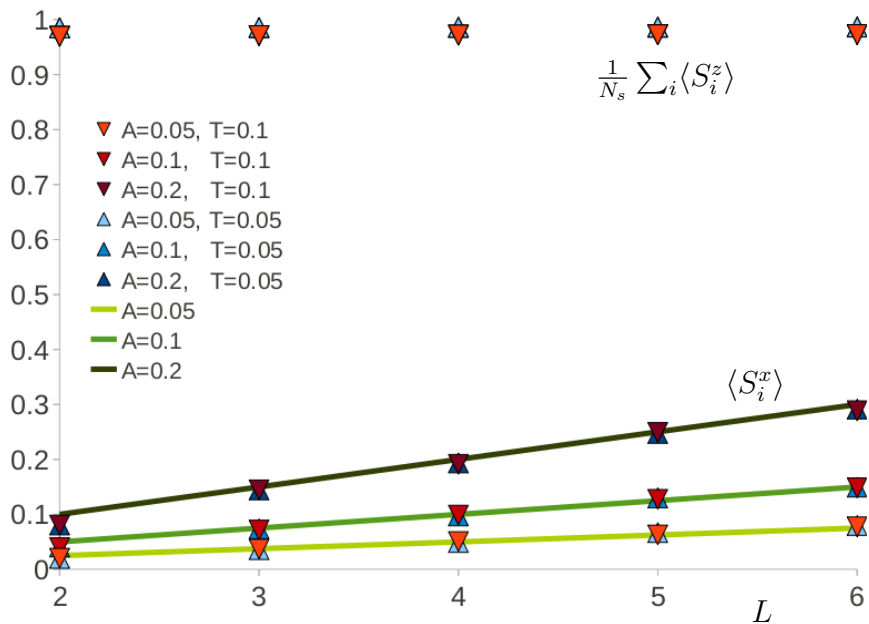

Strictly speaking, our analysis of the toric code coupled to a FM (by means of a perturbative Schrieffer-Wolff transformation) is valid when the FM spins are aligned close to the $z$-direction. 
This is the reason why we introduced the external  field $h_{z}$; it stabilizes the FM against the transverse effective magnetic field induced by the toric code and forbids energetically the turning of the total magnetization.
However, backaction effects are substantial for the FM spins close to the code and it is interesting to study the dynamics of the FM spins in contact to the toric code.
Our study of backaction effects involve both analytical and numerical results and give a good picture of the dynamics of a Heisenberg ferromagnet subject to transverse magnetic field that is localized in a given plane of the lattice.

Let us consider the situation where the coupling of the surface code to the FM is turned on at $t=0$ and let us calculate the dynamics of the $x$-component of a FM spin ${\bf S}_i$ assuming that $W_{j}=+1\,\forall j$. 
At time $t>0$ we have, $ \langle S_{i}^{x}(t)\rangle=\m{Tr}{\rho_F}S_{i}^{x}(t)$,
where $\rho_{F}=e^{-\beta H_{F}}/\m{Tr}e^{-\beta H_{F}}$ 
and $S_{i}^{x}(t)=e^{iH't}S_{i}^{x}e^{-iH't}$, with $H'$ as in Eq.~\eqref{eq:Hprime}.
Here, $i$ in $S_{i}^{x}(t)$ labels an arbitrary site ${\bf R}_i$ of the FM.
The dynamics of $\langle S^x_i(t)\rangle$ can be calculated exactly as $H'$ is exactly diagonalizable, see Appendix~\ref{app:dynamics}.
We find that
\begin{align}\label{eq:xSpin}
 &\langle S^x_i(t)\rangle\nn\\ &= \frac{A}{\pi J}\sum_{p}\frac{C\left(\frac{|{\bf R}_i-{\bf R}_p|}{\sqrt{4\pi JSt}}\right) + S\left(\frac{|{\bf R}_i-{\bf R}_p|}{\sqrt{4\pi JSt}}\right) - 1}{|{\bf R}_i-{\bf R}_p|}
\end{align}
where $C(x)$ and $S(x)$ are the Fresnel integrals.
This expression can be evaluated analytically in several limits (see Appendix~\ref{app:dynamics} for details).

Let us first consider a spin at a FM site ${\bf R}_i$ which is directly adjacent to the toric code.
For a small code ($L<JS/A$) we find the long-time limit $\langle S^x_i(t\ra\infty)\rangle=-AL/J$.
We compare this result with a Metropolis simulation of the classical Heisenberg FM and obtain good agreement, see Fig.~\ref{fig:numerics}. 
On the other hand, consider a code which is assumed to be large enough such that it can be formally extended to infinity.
For this case, we find the finite-time behavior
\begin{align}
 \langle S^x_i(t)\rangle = -4A\sqrt{\frac{St}{\pi J}}
\end{align}
for spins adjacent to the toric code.
The FM spins next to the code adapt to the effective magnetic field in $x$-direction in a diffusive way with diffusion constant $\sim A^2S/J$.
Note that this expression diverges in the long-time limit, which is of course unphysical since $|\langle S^x_i(t)\rangle|$ is bounded by $S$. 
This divergence is an artefact of the harmonic (one-magnon) approximation. We thus trust this approximation at most only for times which are such that $|\langle S^x_i(t)\rangle|\leq S$.

The deviations $\langle S^x_i(t)\rangle$ in $x$-direction become comparable with $S$ after a time of order $t_r\sim JS/A^2$.
We refer to $t_{r}$ as the refreshing time: at this time, the backaction of the surface code on the FM has become substantial with the FM spins close  to the code being tilted away from the magnetization direction of the FM (along the $z$-axis) and pointing now along the $x$-axis. 
To restore the full effect of the FM, we refresh the ferromagnetic state with, e.g., a magnetic pulse, so that all spins point again along the $z$-axis. 
This procedure has to be repeated periodically on a time scale $t_{r}$, which, importantly, is independent of the code size $L$. 
This refreshing can be considered as part of a cooling cycle to get the heat generated by the surface code out of the system (note that no measurements of stabilizers or entangling operations are involved). 
This refreshing prevents the total system, FM plus surface code, to reach a new common equilibrium state, and instead ensures that the FM stays in its own equilibrium state.

Finally, let us consider a spin at a FM site ${\bf R}_i$ with a distance $d$ away from the toric code, which is again assumed to be very large.
For this case, Eq.~\eqref{eq:xSpin} evaluates to 
\begin{align}
 &\langle S^x_i(t)\rangle \nn\\&= \frac{16A}{d^2}\sqrt{\frac{JS^3t^3}{\pi}}\left(\cos\left(\frac{d^2}{8JSt}\right)+\sin\left(\frac{d^2}{8JSt}\right)\right)+O(d^{-3})\,.
\end{align}
The deviation of the FM spins from the ordering along $z$-direction decays quadratically with the distance from the code.

\subsection{Longitudinal coupling to the ferromagnet}\label{sec:longitudinal}

We note that the refreshing process represents a sufficent condition for maintaining an effective quantum memory Hamiltonian. 
However, it is not necessary. 
Indeed, let us consider the extreme case where all the spins of the FM tilt into $x$-direction (possible if we allow $E_{x}$ to exceed $E_{z}$ by assuming e.g. $h_z=0$). 
In this worst case scenario, the interaction between plaquettes is not given by the transverse susceptibility of the FM anymore but by the longitudinal one. 
This fact is derived perturbatively in more detail in Appendix~\ref{app:SWlongitudinal}.
The longitudinal susceptibility of the FM has been studied in detail both with a spin wave analysis \cite{Kawasaki} and with a decoupling method \cite{Callen1, Callen2}. 
The small ${\bf q}$ result reads $\chi_{\vert\vert}({\bf q},\omega=0)=k_BT/8 D^2{\vert{\bf q}\vert}$.
This is valid when $h\ll Dq^2\ll k_BT$, which is the regime of interest here since we focus on distances smaller than $L_{h}$. 
We note that, contrary to the transverse susceptibility,  $\chi_{\vert\vert}({\bf q},\omega=0)$ vanishes at $T=0$, since it corresponds to particle-hole excitations.
Here $h$ points in longitudinal direction and is composed of an external magnetic field (which, as above, is assumed to scale as $1/L^4$) and the magnetic field produced by the surface code. 
Since the latter scales as $L^2/\Lambda^3\propto 1/L^7$, see Appendix~\ref{app:SWlongitudinal}, the longitudinal field produced by the memory can safely be ignored. The magnetic length thus scales again as $L_{h}\propto L^2$. In real space we have
$\chi_{\vert\vert}({\bf r})\propto T/r^2$. 

This worst case is thus analogous to choosing $O_p=a_p\mdag a_p$ in the Hamiltonian discussed in Sec.~\ref{sec:previous}, for which the anyon chemical potential grows logarithmically with $L$ and the memory lifetime grows polynomially.
Again, this is what one would have expected, since in the Holstein-Primakoff picture the longitudinal spin component at ${\bf R}_p$ is given by $-S+a_p\mdag a_p$.

\section{Conclusions}\label{sec:conclusions}

Whether there is a physical Hamiltonian that allows for thermally stable storage of quantum information is a big open question in theoretical physics.
The answer will depend on what properties are required for a Hamiltonian to deserve the label ``physical''.
If one only requires bounded strength and locality of interactions, the answer is affirmative, as the 4D toric code shows \cite{Alicki2010}.
The 4D toric code also constitutes the only known example of a quantum memory Hamiltonian for which it is rigorously proven that the lifetime grows without bounds (below a critical temperature) and that it is stable against arbitrary (local and weak enough) perturbations.
Requiring locality in at most three dimensions excludes the 4D toric code, but boson-mediated long-range interactions still can make the energetic penalty for creating an anyon arbitrarily high \cite{Pedrocchi2011,Hutter2012,Pedrocchi2013}.
If one additionally requires that all operators be bounded (and thereby excludes bosonic operators), the cubic code \cite{Haah2011a,Haah2011b,Bravyi2013} still exhibits self-correcting behavior.
Finally, if one takes into account that interactions in nature are in fact two-body and thus requires that all terms in the Hamiltonian involve at most two spins, one excludes all existing proposals.

In this work, we have shown that even under this most rigid understanding of what constitutes a ``physical'' Hamiltonian, we can still expect to observe self-correcting behavior.
The quantum memory Hamiltonian of Ref.~\onlinecite{Pedrocchi2013} emerges as the low-energy effective theory of a 3D model with bounded-strength interactions between nearest-neighbor spins only.
This effective Hamiltonian describes the low-energy, long-wavelength response of the system, and self-correcting behavior can only be observed if the noise affecting the memory is such that this description remains valid.

\section{Acknowledgements}

This work was supported by the Swiss NF, NCCR QSIT, and IARPA.
F.L.P.\ is grateful for support from the Alexander von Humboldt foundation.

\appendix

\section{Schrieffer-Wolff transformation}\label{app:SW}

Consider some Hamiltonian $H_0$ with an energetic gap $\Delta$ between a low- and a high-energy subspace. In the context of the perturbative gadgets in Sec.~\ref{sec:gadgets}, $H_0$ will simply describe the energy splitting $\Delta$ of a \emph{mediator qubit}, \textit{i.e.}, an auxiliary qubit that mediates interactions between adjacent qubits.
Given some perturbation $V$ with $\norm{V}<\frac{\Delta}{2}$, the modified Hamiltonian $H_0+V$ will display a low-energy subspace of the same dimension as the one of $H_0$, which is separated from the high-energy spetrum by a gap of at least $\Delta-2\norm{V}$.
The SW transformation is then defined as a unitary operator $e^S$ (with $S$ anti-Hermitian), such that $e^S(H_0+V)e^{-S}$ is block-diagonal with respect to the high- and low-energy subspaces of $H_0$.
Together with the requirement that $S$ be block-off-diagonal with respect to the low- and high-energy subspaces of $H_0$ and that $\norm{S}<\frac{\pi}{2}$, this specifies $S$ uniquely \cite{Bravyi2011}.

Let $P$ denote the projector onto the low-energy subspace of $H_0$.
The \emph{effective low-energy Hamiltonian} 
\begin{align}
 \heff = Pe^S(H_0+V)e^{-S}P
\end{align}
can be expanded in a perturbative series $\heff = \heff^{(0)} + \heff^{(1)} + \heff^{(2)} + \heff^{(3)} + \ldots$, where explicit formulae for the low-order effective terms can be derived from Ref.~\onlinecite{Bravyi2011}.
Let $Q=\id-P$, $\vd=PVP+QVQ$, and $\vod=PVQ+QVP$. We have
\begin{align}
 \heff^{(0)} = PH_0P\,,
\end{align}
\begin{align}
 \heff^{(1)} = P\vd P\,,
\end{align}
\begin{align}
 \heff^{(2)} = \frac{1}{2}P\left[L_0^{-1}\vod,\vod\right]P\,,
\end{align}
and
\begin{align}
 \heff^{(3)} = \frac{1}{2}P\left[L_0^{-1}\left[L_0^{-1}\vod,\vd\right],\vod\right]P\,.
\end{align}
In the last two expressions, $L_0$ is the Liouvillian superoperator $L_0O=[H_0,O]$, whose inverse is given by
\begin{align}
 L_0^{-1}O = -i\lim_{\mu\rightarrow0^+}\int_0^\infty\m{d}t\,e^{-\mu t}e^{iH_0t}Oe^{-iH_0t}\,.
\end{align}
For the second order effective Hamiltonian, one finds the concise formula
\begin{align}\label{eq:SW2}
 \heff^{(2)} = -\frac{i}{2}\lim_{\mu\rightarrow0^+}\int_0^\infty\m{d}t\,e^{-\mu t}P\left[\vod(t),\vod\right]P\,.
\end{align}
We note that with $H_0 = -\frac{\Delta}{2}\pz_r$ we have
\begin{align}
 L_0^{-1}\px_r = -\frac{i}{\Delta}\py_r\,,
\end{align}
which leads to Eq.~\eqref{eq:SWfinal}.

\section{Interactions mediated by a translationally invariant system}\label{app:mediated} 

\subsection{Coupling to the transverse component of the FM spins}\label{app:SWtransverse}

We show here a detailed derivation of Eq.~\eqref{eq:effectivecoupling} of the main text with the use of a perturbative Schrieffer-Wolff transformation similar to Ref.~\onlinecite{Braunecker2008}. 

We assume here that the FM is in broken-symmetry state with magnetization along $z$-direction and we couple the surface code to the transverse $x$ component of the FM spins:
\begin{equation}\label{eq:Hamiltonian_supp}
H=H_{0}+V=H_{0}+\sum_{{\bf q}}S_{{\bf q}}^{x}A_{-{\bf q}}\,,
\end{equation}
where $H_{0}$ is a general ${\bf S}$-spin Hamiltonian and $A_{i}$ arbitrary operators which commute with $H_{0}$ and with each other. The Fourier components are defined through $S_{{\bf q}}=\frac{1}{\sqrt{N_{s}}}\sum_{i}e^{-i{\bf q}\cdot{\bf R}_{i}}{\bf S}_{i}$ and $A_{\bf q}=\frac{1}{\sqrt{N_{s}}}\sum_{i}e^{-{\bf q}\cdot{\bf R}_{i}}A_{i}$, where $N_{s}$ denotes the number of spins ${\bf S}_{i}$ and $ {\bf R}_{i}$ their site. Here we identify the projector $P$ as the operator projecting onto the subspace with a fixed number of magnons $n_{\bf q}$. Since $S^x$ does not conserve the number of magnons, it is clear that $V_{\m{d}}=0$ and $V_{\m{od}}=V$. Note that we have absorbed the symmetry-breaking term $h_{z}\sum_{i}S_{i}^z$ into $H_{0}$. 
Up to second order, we obtain from Eq.~\eqref{eq:SW2}
\begin{align}
\heff^{(2)}&=-\frac{i}{2}\lim\limits_{\eta\rightarrow0^{+}}\sum_{{\bf q},{\bf q}'}\int_{0}^{\infty}\m{d}t\,e^{-\eta t}\left[S_{{\bf q}}^{x}(t)A_{-{\bf q}},S_{{\bf q}'}^{x}A_{-{\bf q}'}\right]\nonumber\\
&=-\frac{i}{2}\lim\limits_{\eta\rightarrow0^{+}}\sum_{{\bf q},{\bf q}'}\int_{0}^{\infty}\m{d}t\,e^{-\eta t} \times\nn\\&\quad 
\left([S_{{\bf q}}^{x}(t),S_{{\bf q}'}^{x}]A_{-{\bf q}'}A_{-{\bf q}}+S_{\bf q}^{x}(t)S_{{\bf q}'}^{x}\underbrace{[A_{-{\bf q}},A_{-{\bf q}'}]}_{=0}\right)\,.\nonumber\\
\end{align}
We assume that the ${\bf S}$-spins are in thermal equilibrium, described by the canonical density matrix $\rho=e^{-\beta H_F}/\m{Tr}\,e^{-\beta H_F}$, where $H_F$ is the ${\bf S}$-spin Hamiltonian without the coupling to the plaquettes and corresponds to the main part of the Hamiltonian in Eq.~(\ref{eq:Hamiltonian_supp}), \textit{i.e.}, $H_{F}=H_{0}$. 
In doing so, we neglect the backaction of the toric code on the ferromagnet. 
This backaction will be addressed in Appendix~\ref{app:dynamics} below where we show that it leads to a localized effect on the ferromagnet which becomes relevant when the size of the toric code increases (see also Sec.~\ref{sec:backaction} in the main text).
Here, we rely on a formal perturbation expansion in powers of $\norm{A_i}/J$. Convergence of this formal expansion is an interesting question by itself and can be approached along the  lines discussed in Ref.~\onlinecite{Bravyi2011}. 
However, such rigorous treatment is beyond the present scope. 
Still, in the one-magnon (or harmonic) approximation, 
the effective Hamiltonian Eq.~\eqref{eq:effectivecoupling} in the main text is exact in all orders of $\norm{A_i}$, thus showing that all higher order contributions of the Schrieffer-Wolff expansion vanish exactly in the one-magnon regime.

The  equilibrium expectation values are denoted by $\langle\ldots\rangle$. Since $H_0$ is translationally invariant, such that \ $\langle S^\alpha_{{\bf r}_i}S^\beta_{{\bf r}_j}\rangle = \langle S^\alpha_{{\bf 0}}S^\alpha_{{\bf r}_j-{\bf r}_i}\rangle$, we have $\langle S^\alpha_{\bf q}S^\alpha_{\bf q'}\rangle=\langle S^\alpha_{\bf q}S^\alpha_{-{\bf q}}\rangle\delta_{{\bf q}+{{\bf q}'},{\bf 0}}$, and thus
\begin{align}
 \heff^{(2)}&=-\frac{i}{2}\lim\limits_{\eta\rightarrow0^{+}}\sum_{\bf q}\int_{0}^{\infty}\m{d}t\,e^{-\eta t}\left\langle[S_{{\bf q}}^{x}(t),S_{-{\bf q}}^{x}]\right\rangle A_{{\bf q}}A_{-{\bf q}}\nn\\&=-\frac{1}{2}\sum_{\bf q}A_{-{\bf q}}\chi_{xx}({\bf q})A_{\bf q}\ ,
\end{align}
where $\chi_{xx}(\bf q)$ is the static spin susceptibility.

\subsection{Coupling to the longitudinal component of the FM spins}\label{app:SWlongitudinal}
We are now interested in the case where the surface code is coupled to the longitudinal component of the FM spins:
\begin{equation}
H=H_{0}+V=H_{0}+A\sum_{i}W_{i}S_{i}^z\,,
\end{equation}
where the sum runs over the $L^2$ lattice sites lying in the plane of the surface code. The main part $H_{0}$ is the Hamiltonian of the FM, \textit{i.e.}, $H_0=H_F$, which contains the symmetry-braking term $h_z\sum_{i}S_i^z$.
As above, we identify $P$ as the operator projecting onto the subspace with a fixed number of magnons $n_{\bf q}$. In order to distinguish between the diagonal and off-diagonal parts of the perturbation, it is useful to apply the Holstein-Primakoff transformation in the harmonic approximation (see Eq.~(6) in the main text). Doing so we obtain
\begin{equation}\label{eq:perturbation_z}
V=-SA\sum_{i}W_i+A\sum_{i}W_{i}a_{i}^{\dagger}a_{i}\,.
\end{equation}
In Fourier space Eq.~(\ref{eq:perturbation_z}) reads
\begin{equation}
V=-SA\sum_{i}W_i+\frac{A}{N_{s}}\sum_{i}W_{i}\sum_{{\bf q},{\bf q}'}e^{i{\bf R}_{i}\cdot({\bf q}-{\bf q}')}a_{\bf q}^{\dagger}a_{{\bf q}'}.
\end{equation}
It is now straightforward to distinguish between the diagonal and the off-diagonal part of the perturbation, namely
\begin{eqnarray}
V_{\m{d}}&=&-SA\sum_{i}W_{i}+\frac{A}{N_{s}}\sum_{i}W_{i}\sum_{{\bf q}}a_{\bf q}^{\dagger}a_{{\bf q}}\,,\\
V_{\m{od}}&=&\frac{A}{N_{s}}\sum_{i}W_{i}\sum_{{\bf q}\neq{\bf q}'}e^{i{\bf R}_{i}\cdot({\bf q}-{\bf q}')}a_{\bf q}^{\dagger}a_{{\bf q}'}\,.
\end{eqnarray}
Absorbing $V_{\m{d}}$ into the main part of the Hamiltonian, we rewrite
\begin{equation}
H=H'_{0}+V_{\m{od}}\,,
\end{equation}
with (in the harmonic approximation)
\begin{equation}\label{eq:H0prime}
H'_0=-SA\sum_{i}W_i+\sum_{{\bf q}}\epsilon_{\bf q}n_{\bf q}+\frac{A}{\Lambda^3}L^2\sum_{\bf q}n_{\bf q},
\end{equation}
where, as in the main text $\epsilon_{\bf q}=\omega_{\bf q}+h_z$, we assumed that the surface code is free of anyons, \textit{i.e.}, $W_i=+1$, and we used $N_{s}=\Lambda^3$. 
We see from Eq.~(\ref{eq:H0prime}) that the backaction effect of the surface code increases the gap of the magnons from $h_z$ to $h'_z=h_z+AL^2/\Lambda^3$. However this additional term has no weight in the thermodynamic limit since it scales with $\Lambda^{-3}$. Using the specific choice of scaling from the main text, we have $h_z\propto1/L^4$ while $L^2/\Lambda^3\propto1/L^7$. In the thermodynamic limit the magnetic length is thus just given by the external magnetic field $h_z$
\begin{equation}
L_{h'_z}\rightarrow L_{h_z}\propto L^2\,\,\,\,\,\m{for}\,\,\,\,L\rightarrow\infty\,.
\end{equation}
This allows us to safely conclude that the backaction of the surface code is negligible in this case. 

Using Eq.~\eqref{eq:SW2}, we find
\begin{widetext}
\begin{align}
\heff^{(2)}&=\frac{A^2}{2N_{s}^2}\sum_{i,j}W_iW_j\sum_{{\bf q}\neq {\bf q}', {\bf k}\neq {\bf k}'}\frac{e^{i{\bf R}_{i}\cdot({\bf q}-{\bf q}')+{\bf R}_{j}\cdot({\bf k}-{\bf k}')}}{\epsilon_{\bf q}-\epsilon_{{\bf q}'}}\left[a_{\bf q}^{\dagger}a_{{\bf q}'},a_{{\bf k}}^{\dagger}a_{{\bf k}'}\right]\nonumber\\
&=\frac{A^2}{2N_{s}^2}\sum_{i,j}W_iW_j\sum_{{\bf q}\neq {\bf q}'}\frac{n_{\bf q}-n_{{\bf q}'}}{\epsilon_{\bf q}-\epsilon_{{\bf q}'}}e^{i({\bf q}-{\bf q}')\cdot({\bf R}_{i}-{\bf R}_{j})}\nonumber\\
&=\frac{A^2}{2N_{s}^2}\sum_{i,j}W_iW_j\sum_{{\bf q}',{\bf k}}\frac{n_{{\bf k}+{\bf q}'}-n_{{\bf q}'}}{\epsilon_{{\bf k}+{\bf q}'}-\epsilon_{{\bf q}'}}e^{i{\bf k}\cdot({\bf R}_{i}-{\bf R}_{j})}\nonumber\\
&=-\frac{A^2}{2N_{s}^2}\sum_{i,j}W_{i}W_{j}\sum_{{\bf q},{\bf k}}\frac{e^{\beta(\epsilon_{{\bf k}+{\bf q}}-\epsilon_{{\bf k} })}}{\epsilon_{{\bf k}+{\bf q}}-\epsilon_{{\bf k}}}n_{{\bf k}+{\bf q}}(n_{{\bf k}}+1)e^{i{\bf q}\cdot({\bf R}_{i}-{\bf R}_{j})}\nonumber\\
&=-\frac{A^2}{2N_{s}}\sum_{i,j}W_{i}W_{j}\sum_{\bf k}\chi_{zz}({\bf q},\omega=0)e^{i{\bf q}({\bf R}_{i}-{\bf R}_{j})}\label{eq:long}
\end{align}
\end{widetext}
where the last equality comes from the definition of the susceptibility in Eq.~\eqref{eq:susceptibility} evaluated in the one-magnon approximation.
Following the approach of Ref.~\onlinecite{Kawasaki} assuming that $\beta\epsilon_{{\bf q}+{\bf k}},\beta\epsilon_{\bf q},\beta(\epsilon_{{\bf k}+{\bf q}}-\epsilon_{\bf k})\ll1$, we have that 
\begin{equation}\label{eq:zz}
\chi_{zz}({\bf q},\omega=0)=\frac{k_{B}T}{8 D^2}\frac{1}{\vert{\bf q}\vert}\,\,\,\,\m{for}\,\,\,\vert{\bf q}\vert\rightarrow0,
\end{equation}
where $D=2JS$.
From Eqs.~(\ref{eq:long}) and (\ref{eq:zz}), we finally find a chemical potential for the anyons $\mu\propto k_{B}T\ln(L/2)$ as shown in the main text. We note that the term $-SA\sum_{i}W_{i}$ in $H_{0}'$ leads to an increase of the chemical potential by $2SA$. However, this term does not scale with $L$ and can be neglected for large $L$.

\section{Detailed study of the ferromagnetic spin dynamics under the effective longitudinal magnetic field produced by the toric code}\label{app:dynamics}

For simplicity, we write in this Appendix $H$ instead of $H'$, where $H'$ is as given in Eq.~\eqref{eq:Hprime}.
Furthermore, we are interested in the stable regime of the quantum memory, where topological defects are created on time-scales larger than the time on which FM-spins close to the toric code adjust.
We thus assume $W_i\equiv+1$ for all stabilizers $W_i$.
In the one-magnon approximation, we then have 
\begin{align}
 H = \sqrt{2S}A\sum_p(a_p+a_p\mdag)+\sum\kk\epsilon\kk a\kk\mdag a\kk\,,
\end{align}
with $\epsilon\kk=\omg\kk+h_z$ and $\omg\kk=4JS[3-(\cos(k_x)+\cos(k_y)+\cos(k_z))]\approx2JS{\bf k}^2$.

\subsection{Non-equilibrium response for $S^x$}
We now calculate the time-dependent expectation value of the local $x$-magnetization, defined as
\begin{equation}
 \langle S^x_i(t)\rangle=\tr\lbrace\rho_F S^x_i(t)\rbrace
\end{equation}
where
\begin{equation}
 S^x_i(t)=e^{iHt} S^x_i e^{-iHt}
\label{spin_x_def}
\end{equation}
with
\begin{align}
 \rho_{F}=e^{-H_F/k_BT}/Z_F, \,\, Z_F={\rm Tr} e^{-H_F/k_BT}
\end{align}
and $H_F = \sum\kk\epsilon\kk a\kk\mdag a\kk$ in one-magnon approximation.

We define the polaron transformation \cite{Mahan1990,Pedrocchi2013}
\begin{align}
 \mS = \frac{\sqrt{2S}A}{\sqrt{N_{s}}}\sum_p\sum\kk\frac{1}{\epsilon\kk}(e^{i{\bf k}{\bf R}_p}a\kk-\m{h.c.})
\end{align}
which exactly diagonalizes $H$. 
We have
\begin{align}
 e^{\mS}a_ie^{-\mS} = a_i + [\mS,a_i] = a_i - \frac{\sqrt{2S}A}{N_{s}}\sum_{p}\sum\kk\frac{1}{\eps\kk}e^{i{\bf k}({\bf R}_i-{\bf R}_{p})}
\end{align}
and
\begin{align}
 e^{\mS}a\kk e^{-\mS} = a\kk + [\mS,a\kk] = a\kk - \frac{\sqrt{2S}A}{\sqrt{N_{s}}}\sum_p\frac{1}{\eps\kk}e^{-i{\bf k}{\bf R}_p}
\end{align}
Using these two relations, one easily shows that
\begin{align}
 e^{\mS}He^{-\mS} = H_F + \m{const}\,.
\end{align}

We can thus calculate
\begin{align}
 \langle S^x_i(t)\rangle &= \tr\left\lbrace\rho_Fe^{iHt}S^x_ie^{-iHt}\right\rbrace \nn\\
&= \tr\left\lbrace\rho_Fe^{-\mS}e^{iH_Ft}e^{\mS}S^x_ie^{-\mS}e^{-iH_Ft}e^{\mS}\right\rbrace\,.
\end{align}
In the Holstein-Primakoff picture, $S^x_i=\sqrt{2S}(a_i+a_i\mdag)$, such that
\begin{widetext}
\begin{align}
 \langle S^x_i(t)\rangle 
&= \sqrt{2S}\tr\left\lbrace\rho_Fe^{-\mS}e^{iH_Ft}(a_i+a_i\mdag)e^{-iH_Ft}e^{\mS}\right\rbrace - \frac{4SA}{N_{s}}\sum_{p}\sum\kk\frac{1}{\eps\kk}\cos({\bf k}({\bf R}_i-{\bf R}_{p})) \nn\\
&= \sqrt{2S}\frac{1}{\sqrt{N_{s}}}\sum\kk\tr\left\lbrace\rho_Fe^{-\mS}\left(e^{i{\bf k}{\bf R}_i}e^{-\eps\kk t}a\kk+\m{h.c.}\right)e^{\mS}\right\rbrace - \frac{4SA}{N_{s}}\sum_{p}\sum\kk\frac{1}{\eps\kk}\cos({\bf k}({\bf R}_i-{\bf R}_{p})) \nn\\
&= \frac{2SA}{N_{s}}\sum_p\sum\kk\frac{1}{\eps\kk} e^{i{\bf k}({\bf R}_i-{\bf R}_p)}e^{-\eps\kk t}+\m{c.c.} - \frac{4SA}{N_{s}}\sum_{p}\sum\kk\frac{1}{\eps\kk}\cos({\bf k}({\bf R}_i-{\bf R}_{p})) \nn\\
&=\frac{4SA}{N_{s}}\sum_{p}\sum\kk\frac{1}{\eps\kk}\left(\cos[{\bf k}({\bf R}_i-{\bf R}_p]-\eps\kk t) - \cos[{\bf k}({\bf R}_i-{\bf R}_{p})]\right)
\end{align}
\end{widetext}
In order to further evaluate the sum $\sum\kk$, we go to the continuum limit and replace it by the integral $\frac{N_{s}}{(2\pi)^3}\int\m{d}{\bf k}$.
We also use the small-${\bf k}$/small-$h_z$ expansion of the magnon dispersion, $\eps\kk\approx2JS{\bf k}^2$.
We obtain
\begin{widetext}
\begin{align}
 \langle S^x_i(t)\rangle 
&\approx\frac{2}{(2\pi)^3}\frac{A}{J}\sum_{p}\int\m{d}{\bf k}\frac{1}{{\bf k}^2}\left(\cos[{\bf k}({\bf R}_i-{\bf R}_p]-2JS{\bf k}^2t) - \cos[{\bf k}({\bf R}_i-{\bf R}_{p})]\right) \nn\\
&\approx\frac{A}{\pi J}\sum_{p}\frac{1}{|{\bf R}_i-{\bf R}_p|}\left(C(\frac{|{\bf R}_i-{\bf R}_p|}{\sqrt{4\pi JSt}}) + S(\frac{|{\bf R}_i-{\bf R}_p|}{\sqrt{4\pi JSt}}) - 1\right)\,,
\end{align}
\end{widetext}
where $C(x)=\int_0^x\cos(\frac{\pi}{2}t^2)\m{d}t$ and $S(x)=\int_0^x\sin(\frac{\pi}{2}t^2)\m{d}t$ are the Fresnel integrals.

Let us first study a FM spin adjacent to the code plane.
In the long-time limit $t\ra\infty$, we have, approximating the code by a disk of radius $L/2$ around our spin of interested at ${\bf R}_i$,  
\begin{align}
 \langle S^x_i(t)\rangle &= -\frac{A}{\pi J}\sum_{p}\frac{1}{|{\bf R}_i-{\bf R}_p|} \nn\\
&\approx -\frac{A}{\pi J}\int_0^{L/2}\m{d}R(2\pi R)\frac{1}{R} \nn\\
&= -\frac{AL}{J}\,.
\end{align}
As $|\langle S^x_i(t)\rangle|$ is bounded by $S$, this can of course only be valid for code sizes $L\leq\frac{JS}{A}$.
Let us thus consider the opposite limit of a code which is large enough such that the integral in the integration over all code plaquettes can formally be extended to infinity. Then,
\begin{widetext}
\begin{align}\label{eq:inCode}
 \langle S^x_i(t)\rangle
&= \frac{A}{\pi J}\int_0^\infty\m{d}R(2\pi R)\frac{1}{R}\left(C(\frac{R}{\sqrt{4\pi JSt}}) + S(\frac{R}{\sqrt{4\pi JSt}}) - 1\right) \nn\\
&= -4A\sqrt{\frac{St}{\pi J}}\,.
\end{align}
\end{widetext}
The FM spins next to the code align with the local field in a diffusive way, with diffusion constant $\sim A^2S/J$.
For a large enough code, the evolution of a local FM spin is independent of the code size, as expected.
For a spin with distance $d$ to the code, the integration is more involved and one finds
\begin{widetext}
\begin{align}\label{eq:outOfCode}
 \langle S^x_i(t)\rangle 
&= \frac{A}{\pi J}\int_0^\infty\m{d}R(2\pi R)\frac{1}{\sqrt{d^2+R}}\left(C(\frac{\sqrt{d^2+R}}{\sqrt{4\pi JSt}}) + S(\frac{\sqrt{d^2+R}}{\sqrt{4\pi JSt}}) - 1\right) \nn\\
&= \frac{16A}{d^2}\sqrt{\frac{JS^3t^3}{\pi}}\left(\cos(\frac{d^2}{8JSt})+\sin(\frac{d^2}{8JSt})\right)+O(d^{-3})\,.
\end{align}
\end{widetext}
The deviation decays quadratically with the distance from the code.
The results in Eqs.~\eqref{eq:inCode} and \eqref{eq:outOfCode} both diverge as $t\ra\infty$.
This is an artefact of the harmonic (one-magnon) expansion. We thus trust these results only for times $t$ such that $|\langle S^x_i(t)\rangle|\leq S$.

\end{document}